\documentclass[pdflatex,sn-mathphys-num]{sn-jnl}

\usepackage{graphicx}
\usepackage{amsmath,amssymb,amsfonts}
\usepackage{xcolor}
\usepackage{booktabs}
\usepackage{siunitx}
\usepackage{fancyhdr}
\usepackage{lineno}

\begin{document}
\title{Geometrically necessary boundaries accommodate the residual elastic strain in cold-rolled Fe-3\%Si}

\author[1]{\fnm{Aditya} \sur{Shukla}}\email{aditya.shukla@esrf.fr}

\author[2]{\fnm{Nikolas} \sur{Mavrikakis}}
\email{aditya.shukla@esrf.fr}

\author*[1]{\fnm{Can} \sur{Yildirim}}\email{can.yildirim@esrf.fr}

\affil*[1]{\orgname{European Synchrotron Radiation Facility (ESRF)},
           \orgaddress{\street{71 Avenue des Martyrs},
                       \city{Grenoble},
                       \postcode{38000},
                       \country{France}}}

\affil[2]{\orgname{OCAS NV},
           \orgaddress{\street{Pres. J.F. Kennedylaan 3},
                       \city{Zelzate},
                       \postcode{9060},
                       \country{Belgium}}}

\abstract{%
  The relationship between plastic deformation accommodation structures and residual
  elastic strain fields in deformed metals is poorly understood at the
  intragranular scale, largely because no experimental technique has provided simultaneous, three-dimensional, bulk-sensitive access to both fields at the length scale of dislocation boundaries. Here we use dark-field X-ray microscopy
  (DFXM) to map intragranular misorientation and residual elastic strain
  simultaneously in three dimensions within a grain of 50\%
  cold-rolled Fe~3\%Si alloy. We resolve geometrically
  necessary boundaries (GNBs) and incidental dislocation boundary (IDB)
  cell structures in the bulk non-destructively. Correlating the elastic strain field with the segmented plastically deformed substructure reveals that GNBs act as the primary
  carriers and distributors of long range residual elastic strain. GNBs separate
  subdomains of distinct mean $d$-spacing, 
  across the grain volume. The plastic misorientation associated
  with IDBs and dislocation cells develops within GNB-delimited subdomains
  that carry comparatively similar values of elastic strain. This
  supports a mechanistic picture in which GNBs accommodate nearly all the
  long-range residual elastic strain in the deformed state, while plastic
  slip propagates into GNB interiors to organize into IDB cells with
  similar strain levels. The three-dimensional
  misorientation and strain gradients quantified here provide direct
  experimental input for recovery and recrystallization modelling in ferritic steels, such as
  electrical steels.%
}


\maketitle
\pagestyle{fancy}
\fancyhf{}
\fancyfoot[C]{\footnotesize Under review at \textit{Metallurgical and Materials Transactions A}.}
\renewcommand{\headrulewidth}{0pt}
\renewcommand{\footrulewidth}{0pt}

\section{Introduction}\label{sec:intro}

Plastic deformation of crystalline metals is primarily accommodated by
the collective motion of dislocations, which facilitate shear deformation
by slip between atomic planes. This is a dissipative process in which a
large fraction of the mechanical work is released as heat, while the
remainder is retained in the lattice as stored energy associated with the
elastic stress fields of the accumulated dislocation content \cite{BEVER19735}. During straining, dislocations are continuously created,
move, interact, and become mutually trapped; this accumulation is far from
spatially uniform. Adjacent regions of the same grain deform by different
amounts and on different combinations of slip systems, and the crystal
must maintain mechanical compatibility across these regions. This
requirement gives rise to geometrically necessary dislocations (GNDs),
which accommodate the strain differences between neighboring regions of
the grain and manifest as measurable lattice curvature and intragranular
misorientation gradients \cite{NYE1953153,Ashby01021970}. The grain
orientation spread (GOS), which integrates this accumulated misorientation
over the grain volume, is consequently a key deformation metric: it
quantifies the heterogeneity of the plastic substructure and correlates
directly with the stored energy that drives subsequent thermally-activetaed phenomena such as recovery and recrystallization\cite{humphreys1995recrystallization}. In many engineering materials, such
as silicon steels, the recrystallization texture depends critically on the
spatial distribution of stored energy and orientation gradients inherited
from the deformed state \cite{doi:10.1080/02670836.2016.1231746,Hutchinson1999}.

In materials with high stacking fault energy, such as low-carbon steels
and Fe--Si alloys, dislocations self-organize into a cellular structure
as strain increases \cite{Hansen01082011,LIU19985819,HUGHES2003147}.The resulting microstructure is hierarchical: grains are
subdivided into cell blocks separated by geometrically necessary
boundaries (GNBs), which are formed by ordered dislocation arrays on
specific slip systems and carry a net Burgers vector content whose
associated misorientation increases systematically with strain. Within
the cell blocks, dislocations from multiple slip systems are trapped
statistically to form incidental dislocation boundaries (IDBs), which
enclose dislocation cells with sizes of order 1~\si{\micro\meter} at
deformation levels corresponding to a 50\% rolling reduction \cite{LI20041069,chen_cells_2004}. This hierarchical arrangement controls the
local stored energy landscape and hence the nucleation sites and kinetics
of recrystallization during annealing. Quantitative characterization of
these structures has historically relied on transmission electron
microscopy (TEM), which provides direct imaging of individual dislocations
and boundaries \cite{Hansen01082011} but is restricted to thin foils,
requires extensive destructive preparation, and cannot follow the same
microstructural volume through successive stages of deformation or
annealing \cite{YU20136577}. Electron backscatter diffraction (EBSD)
and high-angular-resolution EBSD extend the accessible field of view and
provide misorientation statistics \cite{WILKINSON2012366,RAABE2002421}, but remain surface-sensitive and two-dimensional, probing a
layer that may not be representative of the bulk in heavily rolled sheet.

Beyond the plastic substructure itself, a second and far less explored field co-exists in the deformed state: the spatial distribution of three-dimensional residual elastic
strains. This field arises from the long-range stress fields of GNB
dislocation arrays, intergranular compatibility stresses, and the
back-stresses of dislocation pile-ups, and it carries the elastic
component of the stored energy whose release drives recovery. Elastic
strain heterogeneity at the scale of individual dislocation boundaries
is expected from dislocation theory: GNBs, with their net Burgers vector
content, should generate long-range elastic fields, whereas IDBs, formed
by statistical trapping with largely cancelling Burgers vectors, should
be elastically screened at the cell scale \cite{HUGHES2003147}.
Yet this prediction has never been tested directly, because no technique
has provided simultaneous, spatially resolved access to both the plastic
substructure and the elastic strain field in the bulk. To our knowledge,
three-dimensional elastic strain fields at the scale of GNBs and
dislocation cells have never been reported in a cold-rolled metal at
industrially relevant deformation levels.

Dark-field X-ray microscopy (DFXM) overcomes these limitations by enabling
non-destructive, full-field imaging of lattice misorientations and elastic
strain within deeply embedded crystalline volumes \cite{simons-2015,poulsen,yildirim_probing_2020}, using an objective lens akin to TEM to
magnify diffracting crystals. Previous DFXM studies have followed
microstructural and strain evolution during early-stage tensile
deformation of aluminium single crystals \cite{zelenika202},
while recrystallization in heavily deformed Fe~3\%Si has been studied in
4D \cite{yildirim-2022,yildirim-2025}. However, a
quantitative description of the deformed microstructure at industrially
relevant rolling reductions remained out of reach until recently, owing
to an improvement in conventional center-of-mass (COM) analysis in heavily
deformed materials \cite{Henningsson2026Darling,Shukla2026DislocationCells} . Recently, it was demonstrated that a multi-peak analysis strategy
implemented in the open-source Python package \texttt{Darling} \cite{Shukla2026DislocationCells,Henningsson2026Darling} resolves individual dislocation cell boundaries
in 50\% cold-rolled Fe~3\%Si that are irrecoverably smeared by the COM
approach, and that a seeded region-growing segmentation extracts cell
size distributions in quantitative agreement with TEM literature values.

In this study, we build on that foundation in two directions. First, we
extend the multi-peak DFXM analysis and dislocation cell segmentation to
three dimensions by acquiring a layer-by-layer volume through a grain
belonging to a well-defined texture component. Second, and centrally, we
exploit the unique capabilities of DFXM to map the residual elastic strain 
within the grain, simultaneously with the orientation field throughout the same
3D volume. By correlating the segmented plastic substructure, GNBs and
IDB cells, with the elastic strain maps, we address a fundamental
question of mesoscale plasticity: do GNBs and IDBs play distinct roles in
accommodating the residual elastic strain? The answer we find is
unambiguous. The residual elastic strain is localized at the grain boundaries and
at the GNBs, while GNB-delimited subdomains carry distinct but internally
homogeneous strain levels, and the IDB cell interiors are regions of
comparatively low residual elastic strain despite carrying substantial plastic
misorientation. 

This
supports a mechanistic picture in which GNBs accommodate nearly all the
long-range residual elastic strain, while plastic slip propagates into
GNB interiors to organize into IDB cells, that is, the cellular
substructure is the product of essentially stress-free plasticity. We
discuss the implications of this separation for the stored energy
landscape and for the recovery and recrystallization behavior of
Fe~3\%Si alloys.

\section{Material and Experimental Methods}\label{sec:methods}

\subsection{Material}\label{subsec:material}

A sample of Fe--3\%Si binary alloy was initially hot rolled and annealed,
resulting in a fully recrystallized ferritic microstructure with an
average grain size of approximately 150~\si{\micro\meter} prior to
deformation. The material was subsequently cold rolled to a true strain
of $\varepsilon_{\mathrm{vm}} \approx 0.8$, corresponding to a 50\%
reduction in thickness. Owing to the high silicon content, the alloy
remained fully ferritic throughout processing. Samples for DFXM were cut
from the rolled sheet and mechanically polished. For reference characterization of the deformed
microstructure, EBSD was performed on a representative specimen from the
same sample batch (not the sample measured by DFXM), using
a JEOL JSM-7001K, Field Emission Gun (FEG) SEM microscope, using a step size of \(0.3~\si{\micro\meter}\) and an accelerating voltage of 20 kV. Phase contrast tomography (PCT) was also measured at ID03 beamline \cite{shukla_bridging_2025} to correlate lab coordinate system with sample coordinate system.  The measured sample had dimensions;  \(1~\mathrm{mm}\) in the rolling direction (RD),\(140~\si{\micro\meter}\) in the transverse direction (TD), and \(200~\si{\micro\meter}\) in the normal direction (ND). The crystallographic texture was mapped at the ID11 beamline of ESRF \cite{WRIGHT2020100818}. The sample was illuminated  with an X-ray beam of 65 KeV while rotating the sample in steps of \(0.1^\circ\). The diffraction data were using a FReLoN camera.

\subsection{Dark-Field X-ray Microscopy}\label{subsec:dfxm}

\begin{figure}[htbp]          
    \centering                
    \includegraphics[width=1\textwidth]{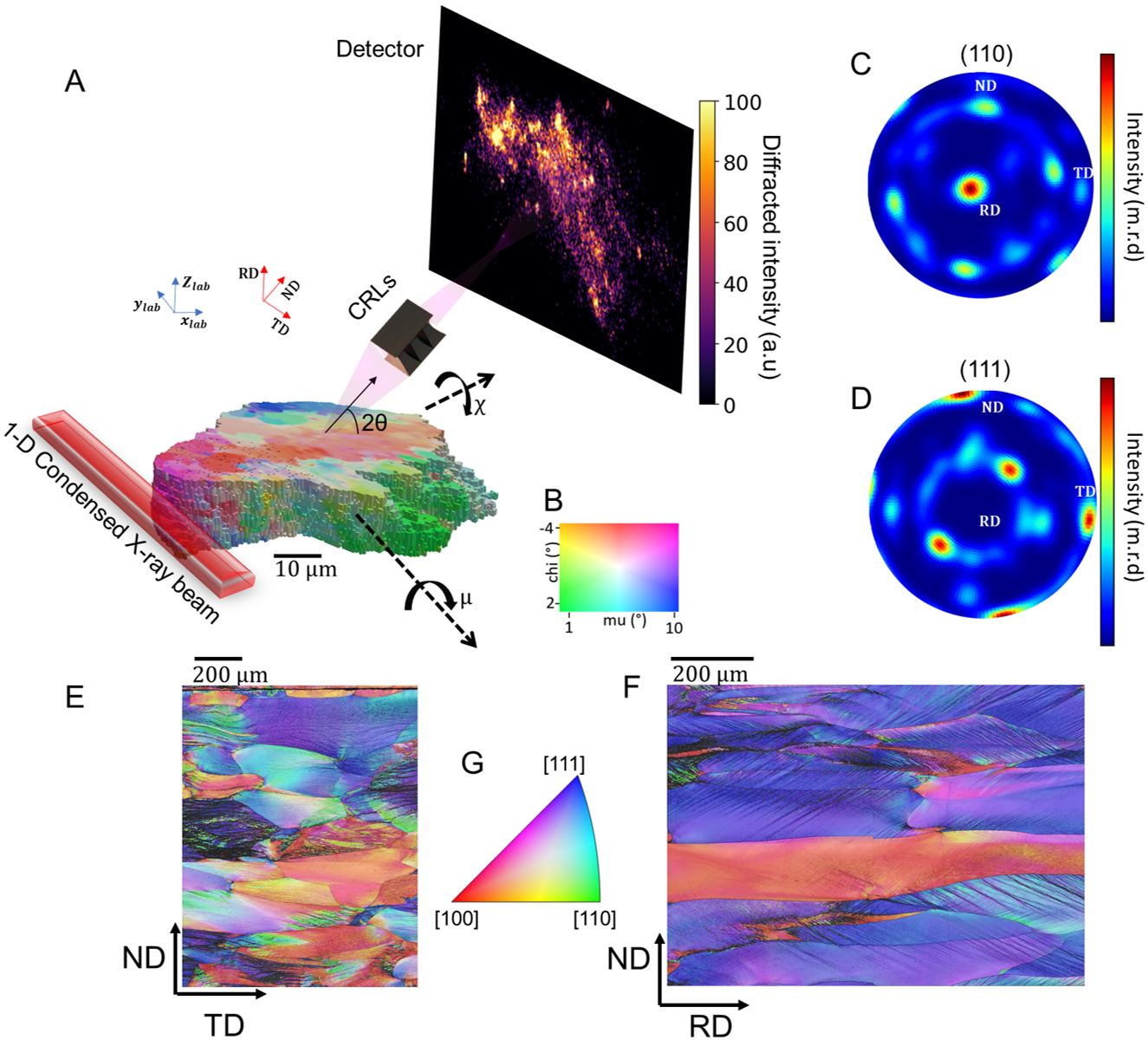} 
    \vspace{-2mm}
    \caption{
    (A) Schematic of the DFXM experimental setup. A one-dimensional focused line beam (500~nm) is used to illuminate the sample. A selected grain is brought into diffraction condition, and the mosaicity maps are reconstructed by scanning the angular motors \(\mu\) and \(\chi\), as indicated in the figure. The laboratory and sample coordinate systems are shown for reference, together with the reconstructed three-dimensional grain volume within the region of interest. The corresponding colour scale is provided (B). (C-D)  Sample averaged pole figures for $\{110\}$ and $\{111\}$ family of planes. (F-G) EBSD maps of the TD-ND and ND-RD cross-sections respectively of a sample which underwent similar processing. (H) IPF-ND color scale for EBSD plots in E and F.}
    \label{fig:setup} 
\end{figure}

DFXM experiments were conducted at the ID03 beamline \cite{id03} of the European Synchrotron Radiation Facility (ESRF). A schematic
of the experimental geometry is shown in Figure~\ref{fig:setup}a. The sample was mounted such that \(\mathrm{RD} \parallel z_{\mathrm{lab}}\). From the PCT measurements , the angle between ND and the laboratory \(Y\)-axis was determined to be \(58^\circ\), while the angle between TD and the laboratory \(X\)-axis was \(32^\circ\).  A monochromatic
X-ray beam with an energy of 17~keV was focused into a line with a full
width at half maximum (FWHM) of approximately 500~nm along
$z_{\mathrm{lab}}$, illuminating a thin layer in the sample. The
measurements targeted the \{110\} reflection of ferritic iron. The
diffracted signal was magnified using a combination of compound
refractive lenses (CRLs) and optical elements of the detector, which was
positioned 5~m downstream of the sample, yielding an effective pixel
size of 160~nm in the $y_{\mathrm{lab}}$ direction. A grain belonging to
the $\langle110\rangle\!\parallel\!\mathrm{RD}$, $\alpha$-fibre, with a
size of approximately 50~\si{\micro\meter}, was selected for detailed
characterization. Local misorientation around the $\langle110\rangle$
direction was measured by scanning the sample over the tilt angles
$\chi$ and $\mu$. Three-dimensional spatial information for of the grain  was obtained by translating the sample along the \(z_{\mathrm{lab}}\) direction in increments of 1~\si{\micro\meter}. At each step, positions, a full \((\chi,\mu)\) scan was recorded, enabling reconstruction of the region of interest in the grain volume on a layer-by-layer basis. To probe the residual elastic strain state, the channel-cut monochromator (CCM) was rocked during acquisition of the central layer, thereby varying the incident X-ray energy. This introduces diffraction contrast associated with local variations in the \(110\) lattice-plane spacing. These measurements provide a measure of the residual elastic strain distribution within the sample.~\cite{id03}. Elastic strain was evaluated from using a reference \(d_{110}\) lattice spacing of \(2.0251~\text{\AA}\), from which the local residual elastic strain was determined.

Previous studies have demonstrated that a multi-peak center-of-mass (COM) analysis is necessary for accurately characterizing DFXM data acquired from deformed materials. Therefore, the data were analysed using the multi-peak COM framework implemented in the Python package \texttt{DARLING}~\cite{Shukla2026DislocationCells,Henningsson2026Darling}.

\subsection{Segmentation of dislocation boundaries}
\label{subsec:segmentation}

Dislocation cells were segmented using an iterative seeded region-growing (flood-fill) algorithm~\cite{adams_seeded_1994}, extended here from the two-dimensional implementation of~\cite{Shukla2026DislocationCells} to the full three-dimensional volume. Seed points were identified in regions of low kernel average misorientation (KAM), ensuring that region growth originates from cell interiors. From these seeds, regions were iteratively expanded subject to two constraints: (i) a KAM-based threshold to prevent crossing of high-angle boundaries, and (ii) an orientation-space distance criterion between neighboring voxels. The segmentation was performed in successive KAM tolerance levels of $\kappa_{\mathrm{KAM}} = 0.1^\circ$, $0.4^\circ$, $0.8^\circ$, and $1.0^\circ$. In addition, Euclidean distance thresholds in the \(\mu\)–\(\chi\) center-of-mass (COM) space were set to 0.1 for region growth and 0.2 for boundary refinement. Further details of the segmentation approach for DFXM mosaicity datasets can be found in~\cite{Shukla2026DislocationCells}. GNBs were identified as extended, planar walls of high misorientation spanning multiple layers of the reconstructed volume. These features were distinguished from IDBs based on their higher misorientation levels and persistent, planar morphology~\cite{HUGHES2003147}.
\section{Results}\label{sec:results}

\subsection{Misorientation and plastic strain}
            \label{subsec:overview}

As discussed above, plastic deformation generates GNDs, which induce local lattice rotations that can be quantified as misorientation. As plastic strain accumulates, the local misorientation increases; 
metrics such as KAM and GOS are therefore widely used proxies for 
spatial variations in plastic strain~\cite{Wright2011,Pantleon2008,KAMAYA201256}. Figure~\ref{fig:setup}F and Figure~\ref{fig:setup}G show EBSD orientation maps of TD-ND and ND-RD cross sections for a 50\% cold-rolled sheet colored using IPF-ND colorscale. IPF-RD EBSD plots for the same region are provided in supplementary information Figure 1. The microstructure exhibits typical deformation features at this reduction, including elongated grains aligned with the rolling direction (RD), pronounced intragranular orientation gradients, and deformation banding. The TD-ND section of the EBSD map has equiaxial grain shape. Figure~1C and Figure~1D show the average texture of the studied specimen. 
It reveals a dominance of the $\alpha$-fibre, 
\(\langle110\rangle \parallel \mathrm{RD}\), consistent with findings reported in similar materials~\cite{MAVRIKAKIS201992} and from our EBSD measurements. Figure~1a shows the reconstructed three-dimensional volume of a selected \(\langle110\rangle \parallel \mathrm{RD}\), \(\alpha\)-fiber grain measured by DFXM, visualized in terms of the mosaicity (orientation spread) field. The grain exhibits an average orientation spread of approximately \(5^\circ\), consistent with previous studies~\cite{MAVRIKAKIS201992}. The TD-ND crossection of the 3D DFXM grain volume shown in Figures~\ref{fig:setup}A (TD-ND cross-section is in the same plane as the laboratory $x$--$y$ plane) is measured to be equiaxed as expected from EBSD measurements.

\begin{figure}[htbp]          
    \centering                
    \includegraphics[width=1\textwidth]{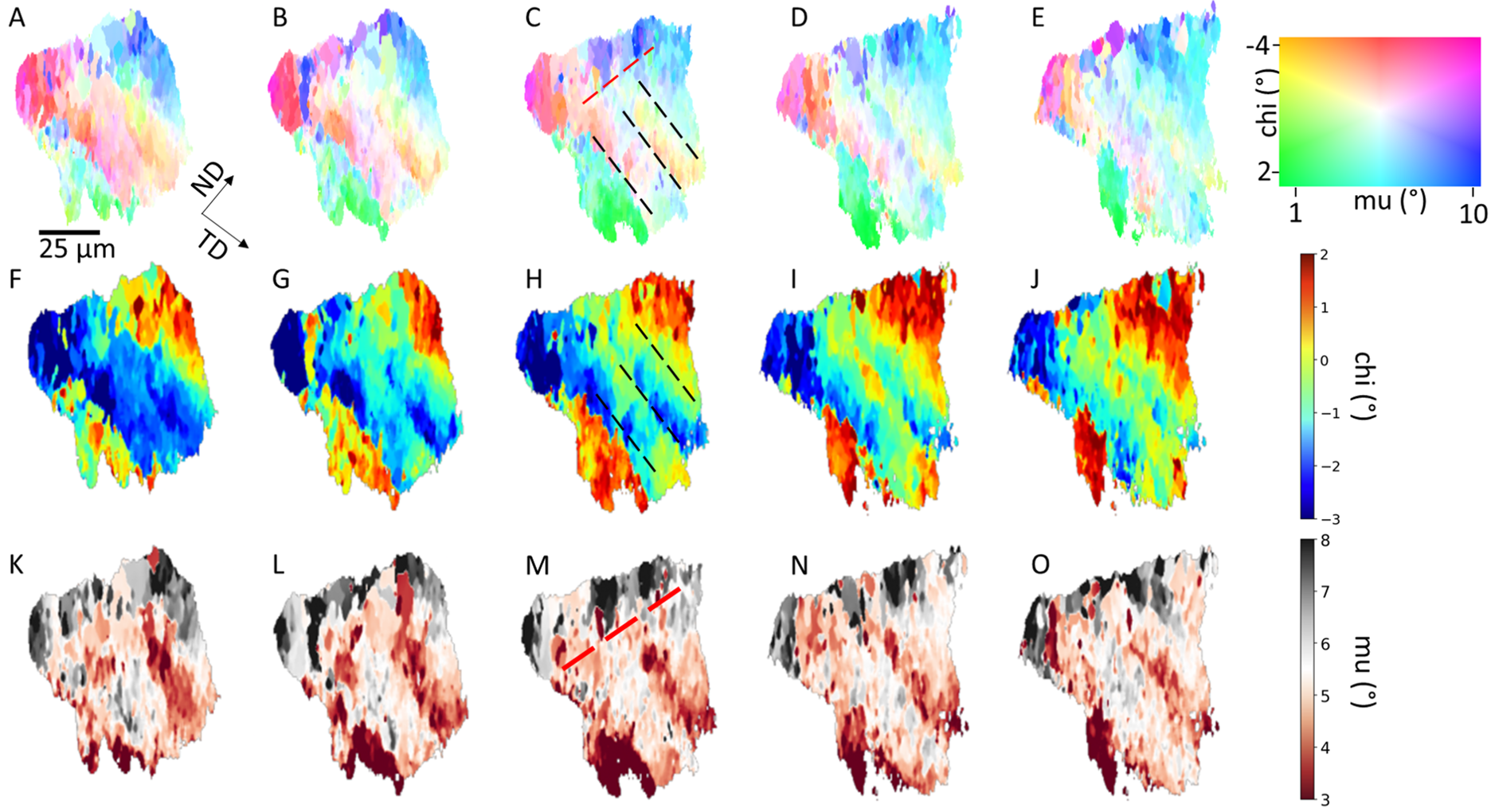} 
    \vspace{-2mm}
    \caption{
    (A–E) Mosaicity maps corresponding to TD–ND cross-sections reconstructed at successive positions along the rolling direction (RD), with a spacing of \(2~\si{\micro\meter}\) between slices. (F–J) \(\chi\) center-of-mass (COM) maps for the same regions. (K–G) \(\mu\)-COM maps illustrating spatial variations in local orientation. Black and red lines serve as guides to the eye, indicating the positions of geometrically necessary boundaries (GNBs).
    }
    \label{fig:Mosaicity} 
\end{figure}

Figures~\ref{fig:Mosaicity}A--E show the reconstructed mosaicity maps of the grain of interest for each TD-ND cross-section. The color scale indicates a total intragranular misorientation of approximately \(6^\circ\), consistent across all layers along the rolling direction (RD). Details of the mosaicity color reconstruction are provided in~\cite{darfix}. GNBs, are visible as parallel lines delineating regions with different mosaicity colors. These are highlighted by black and red dashed lines to guide the eye.

To resolve the directional character of the deformation, the orientation field is decomposed into rotations about the laboratory axes. Figures~\ref{fig:Mosaicity}F--J show the \(\chi\) component (rotation about the laboratory \(X\)-axis), while Figures~\ref{fig:Mosaicity}K--O show the \(\mu\) component (rotation about the laboratory \(Y\)-axis). As the normal direction (ND) is inclined by \(58^\circ\) to the laboratory \(Y\)-axis and the transverse direction (TD) by \(32^\circ\), the \(\chi\) maps are more sensitive to rotations associated with rotation axis as ND, whereas the \(\mu\) maps are more sensitive to rotations associated with rotation axis as TD.

The \(\chi\) maps exhibit a smooth orientation gradient (notice the change from blue at bottom left in \(\chi\) maps to red at top right), whereas a similar trend is not observed in the \(\mu\) component. This indicates that the dominant lattice bending of the grain is associated with the rotation around ND axis of the sample. In addition, several GNBs are visible in the \(\chi\) maps which are believed to give rise to this orientation gradient across the ND direction of the grain. Interestingly, some GNBs , like the one highlighted in red in Figure~\ref{fig:Mosaicity}C, appear in the \(\mu\) maps but not in the \(\chi\) maps. These suggests that GNBs have a well defined orientation axis for the misorientation of the lattice. This observation is consitent with earlier work on GNBs ~\cite{RAABE2002421} that suggests that the misorientation magnitude and axis across GNBs can depend on various factors like activated slip systems, grain neighborhood and parent grain orientation.

\begin{figure}[htbp]          
    \centering                
    \includegraphics[width=1\textwidth]{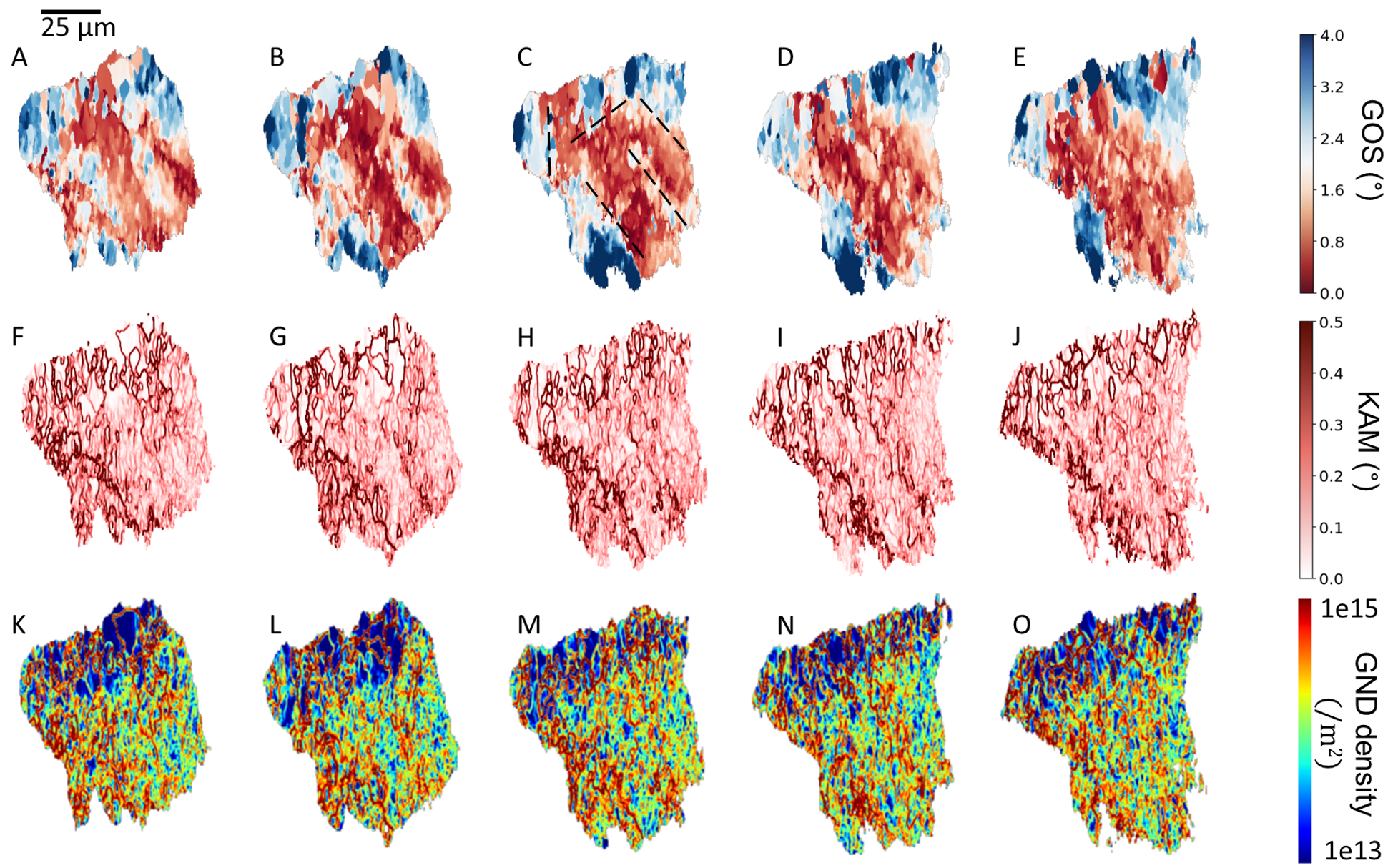} 
    \vspace{-2mm}
    \caption{Layer-resolved misorientation maps across five  slices of the TD-ND plane. (A--E) Grain orientation spread (GOS) maps showing intragranular orientation spread relative to the grain-averaged orientation. (F--J) Kernel average misorientation (KAM) maps. (K--O) GND density maps. GNBs are highlighted as black lines. }
    \label{fig:kam} 
\end{figure}

Figures~\ref{fig:kam}A--E and F--J present layer-resolved GOS and KAM maps, respectively. GOS quantifies the overall intragranular orientation spread relative to the grain-averaged orientation:
\begin{equation}
\mathrm{GOS} = \sqrt{(\chi - \langle \chi \rangle)^2 + (\mu - \langle \mu \rangle)^2},
\label{eq:GOS}
\end{equation}
where $\chi$ and $\mu$ are local orientation components of each voxel and $\langle \chi \rangle$, $\langle \mu \rangle$ their grain-averaged counterparts. KAM provides a local misorientation measure computed as the average misorientation over a $3\times3$ kernel:
\begin{equation}
\mathrm{KAM}_i = \frac{1}{M} \sum_{j=1}^{M} \Delta\theta_{ij},
\label{eq:KAM}
\end{equation}
where $\Delta\theta_{ij}$ is the misorientation between voxel $i$ and neighbour $j$, and $M$ is the number of neighbours. Further details on KAM computation from DFXM data are given elsewhere~\cite{zelenika202}.

The GOS maps reveal a mean intragranular misorientation of approximately $\sim5^\circ$. GNBs appear as sharp, localised changes in GOS, consistent with the well known nature of GNBs~\cite{HUGHES2003147,LIU19985819}. The KAM maps provide a complementary, finer scale view. GNBs identified in Fig.~\ref{fig:Mosaicity} coincide with elevated KAM. Within the grain interiors, a secondary network of higher KAM boundaries is visible. This is attributed to IDBs that partition the microstructure into dislocation cells. Taken together, the GOS and KAM maps paint a consistent picture of a hierarchical deformation structure observed in earlier TEM work, where GNBs organize the coarse-scale strain field and IDBs define the finer cell wall substructure. Using the segmentation workflow described in the methods section above, high KAM boundaries were segmented into dislocation cells as shown in Fig.~\ref{fig:strain}B. Figures~\ref{fig:kam}K--O are the dislocation density maps calculated using the KAM maps using a similar approach as detailed in other dfxm studies \cite{Sanna2026-vj}. The median dislocation density of $\tilde{\rho} \approx 10^{14}\ \mathrm{m}^{-2}$ measured  is in good agreement with values reported in the literature for similar materials
\cite{MAVRIKAKIS201992,met14101127}. 

\begin{figure}[htbp]          
    \centering                
    \includegraphics[width=1\textwidth]{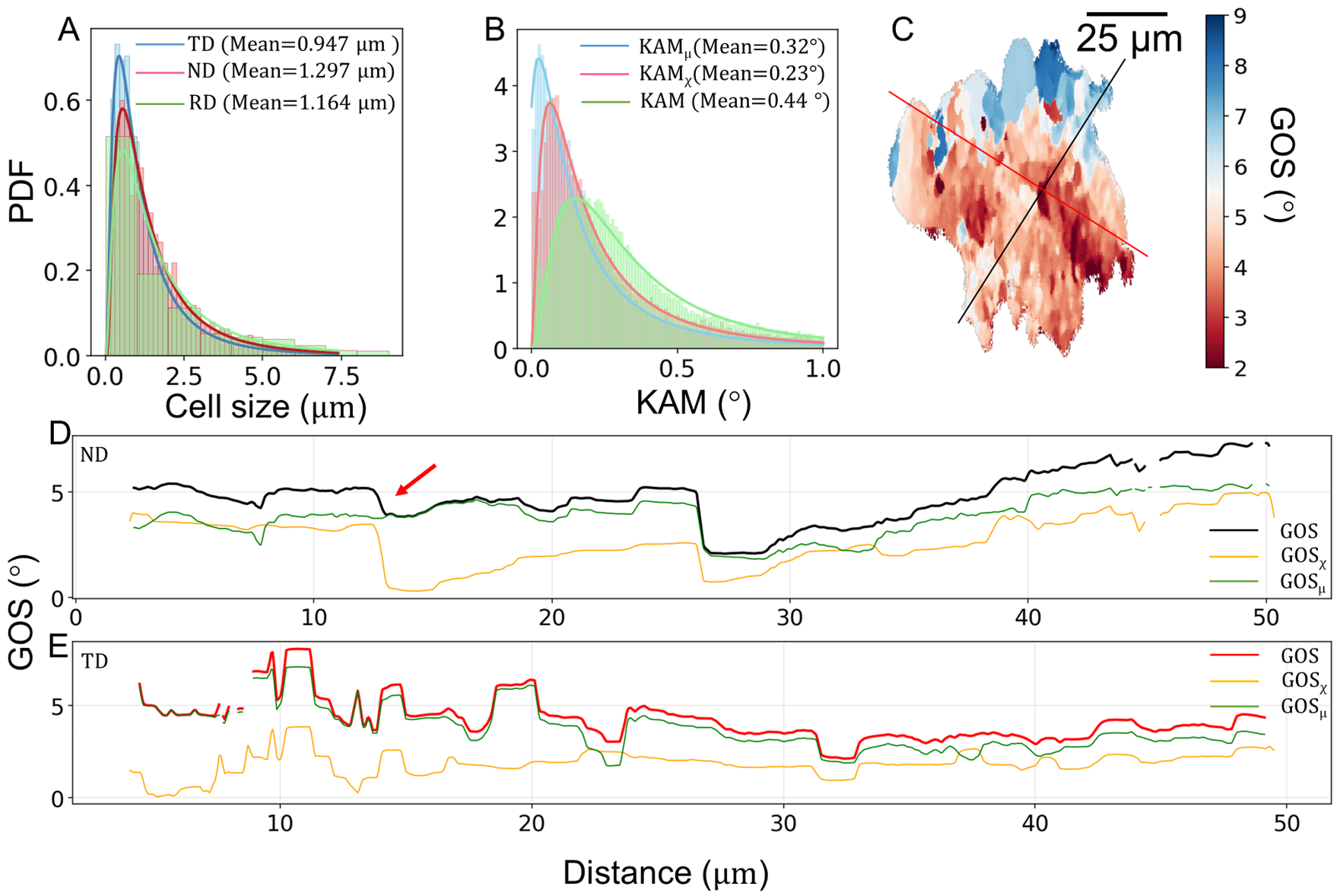} 
    \vspace{-2mm}
    \caption{(A) Probability distribution of dislocation cell sizes measured along TD, ND, and RD (B) KAM probability distributions computed for rotations about $\chi$, $\mu$, and their combination(C) Central TD--ND cross-section of the grain indicating the positions of the line profiles extracted in (D) and (E). (D) GOS line profile along ND, (E) GOS line profile along TD.}
    \label{fig:cell_size} 
\end{figure}

Figure~\ref{fig:cell_size}A shows the probability density distribution of dislocation cell sizes measured along the transverse (TD), normal (ND), and rolling (RD) directions, following the segmentation procedure described above in section ~\ref{sec:methods}. The mean cell size is approximately $1\,\mu$m across all three directions, consistent with values reported for similarly deformed metals in the literature~\cite{YU20136577,chen_cells_2004}. Notably, cells are measured to be more elongated along ND relative to TD by a small amount, indicating that the boundary spacing of the measured grain is smaller in TD than in ND direction.

Figure~\ref{fig:cell_size}B presents the KAM distributions computed separately for rotations about $\chi$ (ND) (ignoring the contribution of $\mu$ in equation~ \ref{eq:KAM} in this case) and $\mu$ (TD), as well as the combined KAM. The mean KAM is $\sim0.4^\circ$, in agreement with values reported for comparable deformation levels~\cite{GODFREY20001897}. Interestingly, we observed a directional dependence on calculated KAM. KAM calculated by only using the contribution of misorientation about $\mu$ (TD) exceeds KAM about $\chi$ (ND), indicating that local orientation changes arising from rotations about TD are larger than those about ND. This is consistent with the $\chi$ maps in Fig.~\ref{fig:Mosaicity}B, where the orientation variation measured along the axis $\chi$ (with ND as the primary misorientation axis) manifests primarily as a smooth, long-range gradient attributable to GNBs, rather than sharp local fluctuations. In contrast, rotations about $\mu$ produce stronger short-range misorientation, reflecting a more heterogeneous local dislocation structure in that direction. 
To investigate the directional dependence of the orientation gradients in greater detail, line profiles of GOS were extracted along both TD and ND across the central TD-ND cross-section of the grain. Profiles along ND exhibit fewer number of discontinuous jumps in GOS value, suggesting that  misorientation along ND is mostly due to the presence of GNBs, whereas profiles along TD reveal a more heterogeneous, more frequent variations indicative of the IDB cell substructure.

To further isolate the rotational contributions, the GOS line profile was decomposed into its $\chi$ (neglecting the contribution of $\mu$ for calculation of GOS in equation ~\ref{eq:GOS}) and $\mu$ (rotation about TD) components. The decomposed maps reveal that certain GNBs, such as the one highlighted with a red arrow in Fig.~\ref{fig:cell_size}, are visible exclusively in the $\chi$ map and absent in the $\mu$ map, demonstrating that these boundaries carry misorientation predominantly through rotation about ND. Conversely, the TD line profile closely follows the $\mu$ map, suggesting that orientation gradients in the TD direction are governed primarily by rotations about TD. The mean gradient of these profiles is 0.26~\si{\degree\per\micro\meter} along ND and 0.295~\si{\degree\per\micro\meter} along TD, the steeper gradient along TD being consistent with the IDB-dominated heterogeneity resolved in that direction. The corresponding profile along RD, spanning the ten reconstructed layers (\(10~\si{\micro\meter}\)), is shown in Supplementary Figure~2: the layer-averaged misorientation increases continuously along RD, carried predominantly by the \(\chi\) component (the one most sensitive to rotation about ND).  Taken together, these observations indicate that misorientation along ND is dominated by rotations about ND, while misorientation along TD is dominated by rotations about TD. This behavior is consistent with the classical framework of geometrically necessary dislocations~\cite{NYE1953153,Ashby01021970}, in which lattice curvature in a given direction is accommodated by GNDs whose Burgers vectors are oriented to produce the corresponding rotation. The preferential activation of specific slip systems during rolling naturally partitions the dislocation content into directionally distinct populations, giving rise to the observed anisotropy in both cell morphology and orientation gradient character~\cite{HUGHES2003147, RAABE2002421}.

\subsection{Residual lastic Strain and its Correlation with the Local Misorientation}\label{subsec:strain}

\begin{figure}[htbp]          
    \centering                
    \includegraphics[width=1\textwidth]{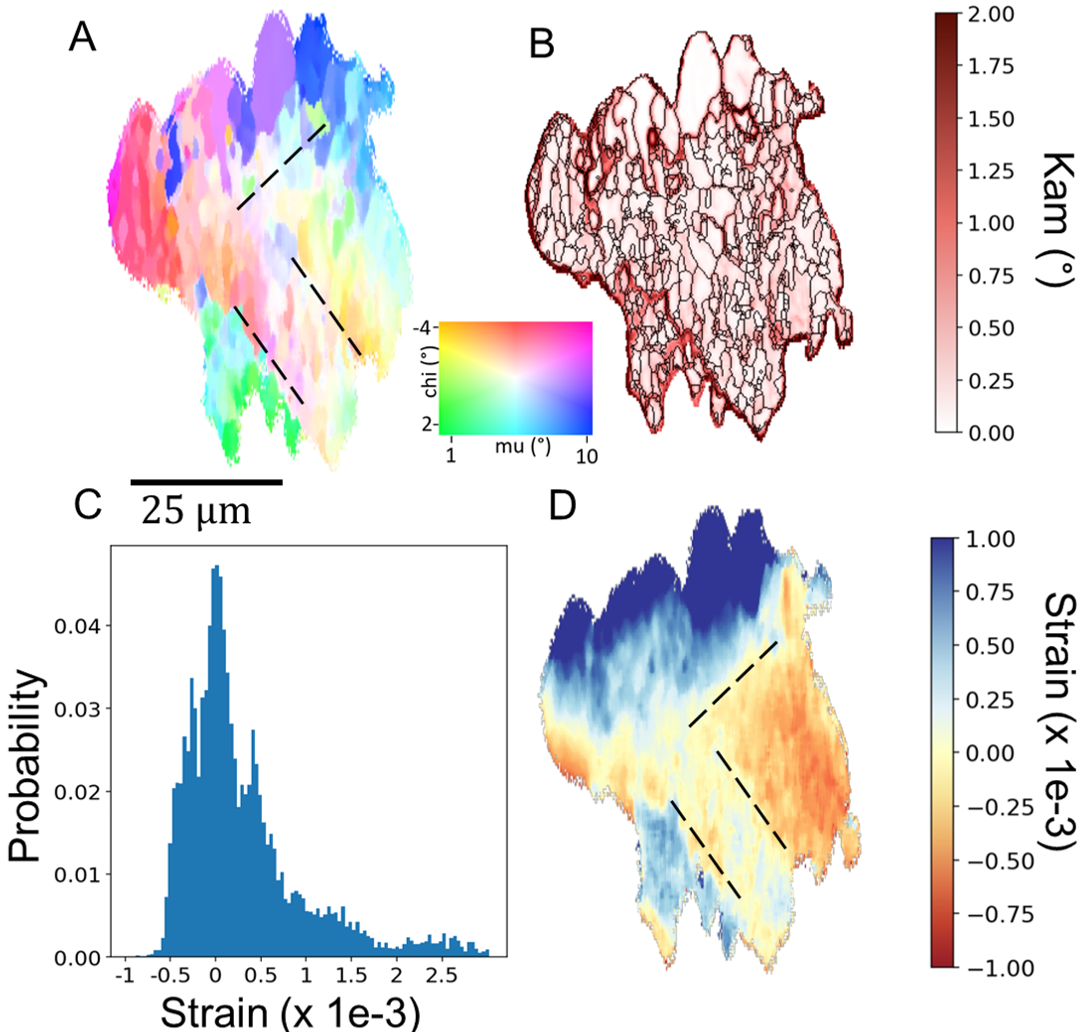} 
    \vspace{-2mm}
    \caption{(A) Mosaicity map of the TD--ND mid-layer cross-section with 
    GNBs highlighted. (B) KAM map of the same layer with dislocation cell 
    boundary segmentation overlaid, highlighting the IDB network. (C) 
    Probability density function of $\varepsilon_{110}$ over the grain 
    volume. (D) Residual elastic strain field ($\varepsilon_{110}$) of the 
    same layer, revealing strain partitioning by the GNB network.}
    \label{fig:strain} 
\end{figure}

The results presented thus far characterize the deformed microstructure 
through the conventional descriptors of orientation mapping, kernel average 
misorientation, boundary classification, and cell statistics, quantities 
that, while obtained here entirely by non-destructive X-ray diffraction techniques, are  
consistent with observations reported in the electron microscopy literature {\cite{LI20041069}}. 
Figure~\ref{fig:strain} presents the bulk residual elastic 
strain field resolved at the scale of the dislocation substructure, measured 
simultaneously with the local misorientation field and spatially correlated 
against it. To our knowledge, such a direct correlation between the 
intragranular strain and misorientation fields at this length scale has not 
previously been reported.

Figure~\ref{fig:strain}A and B show the mosaicity and KAM maps of the 
mid-layer of the three-dimensional region of interest, with GNBs and IDBs 
highlighted respectively. Together, the two panels resolve the full 
dislocation boundary hierarchy within the same measurement volume. Figure ~\ref{fig:strain}D presents the residual elastic strain field expressed as the axial 
strain along the $\langle110\rangle$ direction, equivalently the relative 
variation in $d_{110}$ lattice spacing. There is a spatial correlation
between the GNB network and strain localisation: GNBs partition the grain 
into subdomains of distinct residual strain, while IDBs show no comparable 
effect, consistent with their lower-energy character reported in the 
literature \cite{RAABE2002421,HUGHES2003147}. To our knowledge, this is the first direct experimental 
evidence of this strain-partitioning role, which has remained inaccessible 
due to the multiscale nature of the measurement required to solve this problem.

Next, we quantify grain-averaged elastic strain along 110 direction. Figure~\ref{fig:strain}c shows the strain distribution as a probability density 
function. The width of the histogram is on the order of $\varepsilon_{110} \approx 10^{-3}$, typical in cold-rolled materials. A tail of elevated tensile strain values is also 
visible. From figure~\ref{fig:strain}D, this region of high tensile strain or large d-spacing comes from the top surface of the grain near the grain boundary.

\section{Discussion}
The central experimental finding of this study is that the distribution  of long range residual elastic strains is governed by the distribution of GNBs  while the incidental
dislocation cell interiors have comparatively similar elastic strain values
(Figure~\ref{fig:strain}). This observation brings a direct interpretation in terms of the
distinct dislocation character of the two boundary classes, GNBs and IDBs~\cite{MUGHRABI19831367}. GNBs form by the organized
accumulation of dislocations on a small number of geometrically selected
slip systems and therefore carry a net Burgers vector
content~\cite{HUGHES2003147}. This net content is not screened at the local
scale, in close analogy with the elastic fields of low-angle grain
boundaries~\cite{PhysRev.78.275}, it generates long-range stresses that
extend over distances comparable to the cell-block width. It is precisely
these fields that DFXM resolves as extended regions of elevated
$\varepsilon_{110}$ seperated by GNBs. Incidental
boundaries, by contrast, form through the statistical trapping of
dislocations drawn from several slip systems whose Burgers vectors largely
cancel; their stress fields are screened within a few cell diameters,
leaving the enclosed volumes with similar elastic strain values.  Figure~\ref{fig:strain} reveals elastic residual strains differences reaching ${\sim}5\times10^{-4}$ between the two sides of the GNB. The low elastic strain differences measured
within the cell blocks is the spatially-resolved experimental evidence of
this screening, shown here in the bulk of a 3D-specimen.
Previous experiments inferred this only indirectly from misorientation statistics
and scaling relations~\cite{HUGHES2003147,Hansen01082011}.
 
The finding that adjacent GNB-delimited subdomains carry
distinct but internally homogeneous elastic strains is a real-space image of
the long-range internal stresses whose existence, magnitude, and even
reality have been debated for decades on the basis of indirect and
volume-averaged probes~\cite{KASSNER201344}. Each cell block accomodates
a different combination and amount of slip, set by its own active systems
and by the constraints imposed by neighbouring blocks and
grains~\cite{RAABE2002421}, and therefore carries a distinct internal
back-stress. The contrast in mean $\varepsilon_{110}$ across a GNB provides a
direct, local estimate of that back-stress,
$\sigma_{\mathrm{b}} \approx E\,\Delta\varepsilon_{110}$, (where $E$ is the Young's modulus along the $\langle110\rangle$ direction)
of the same kind that kinematic-hardening laws invoke to reproduce the
Bauschinger effect but rarely measured experimentally at this scale. 

It must be noted that the first-order (mean) elastic strain field, although
it resolves long-range internal stresses with unprecedented spatial
resolution, does not directly represent the driving force for recovery.
The dominant contribution to the stored energy resides in the dislocation
density, but also the second-order (variance) of the local strain
distribution (analogous to peak width in powder XRD measurements)  along with in its mean (analogous to peak position in powder XRD measurements). This second-order strain quantity can
be interpreted as the spread of lattice spacings within a single diffracting gauge
volume in the DFXM experiment. The variance of $d_{110}$ spacings across the measured
grain is presented in Supplementary Information Figure~3. The relatively
homogeneous distribution of second order lattice spacings observed throughout the grain
suggests that the the stored energy due to local elastic strain fields, is
spatially uniform. Hence,the cell interiors are not stress-free, but rather free of long-range internal stress while retaining a dense
statistically stored population and its short-range fields.

The directional character of the misorientation field carries further
crystallographic information accessible through the bulk three dimensional measurements. The KAM computed about $\mu$ is larger
than that computed about $\chi$, and the two components differ in spatial
character: the $\chi$ field, which is more sensitive of to rotations
about ND, is dominated by a smooth long-range gradient, whereas the $\mu$
field, more sensitive to rotations about TD, shows stronger short-range
fluctuation. Because ND and TD are inclined to the laboratory axes (by
$58^\circ$ and $32^\circ$ respectively), $\chi$ and $\mu$ COM maps are a linear combination of misorientations measured about TD and ND.
Hence this correspondence is best read as a tendency rather than a strict
decomposition. With that caveat, the smooth $\chi$ gradient is plausibly
associated with the GNBs and the $\mu$ fluctuation with the IDB cell
structure, consistent with the boundary classification of the preceding
sections. Moreover, some GNBs, as shown in Figs.~\ref{fig:Mosaicity} and
\ref{fig:cell_size}, appear in only one of the DFXM COM maps, implying
that the corresponding lattice misorientation gradient occurs about a misorientation
axis perpendicular to the DFXM rotation axis that cannot measure this misorientation. This provides direct
three-dimensional experimental evidence, obtained from a bulk specimen,
that GNBs are characterized by specific crystallographic misorientation
axes, consistent with the established theory of geometrically necessary
boundaries.~\cite{RAABE2002421} For a grain of the $\langle110\rangle\!\parallel\!\mathrm{RD}$,
$\alpha$-fiber the orientation places strong constraints on the Schmid
factors of the $\{110\}\langle111\rangle$ and $\{112\}\langle111\rangle$ slip
systems~\cite{doi:10.1080/02670836.2016.1231746}, which raises the
possibility that the measured rotation axes could be related to specific
Burgers-vector populations. Establishing such a relationship quantitatively
would require the full lattice-curvature tensor and lies beyond the scope of this work. We note it here as a potential use case of such
three-dimensional data. In the same way, reconstructing the boundaries across successive layers would allow their habit planes to be extracted and compared with slip-plane traces and with the classical lamellar and microband geometries of rolled BCC metals~\cite{HUGHES2003147,RAABE2002421}. Furthermore, GNBs visible in the $\chi$ maps but absent in the $\mu$ maps would be candidates for boundaries of predominantly tilt character with a well-defined misorientation axis, although this interpretation cannot be established from the present data alone.
 
 
These observations have important implications for the mechanisms of recovery and recrystallization which will occur upon subsequent annealing and are governed by stored energy.
A quantitative accounting of the stored energy is presented. Three stored energy reservoirs must be
distinguished. The first is the long-range elastic strain energy carried by the GNB fields, of order $u_{\mathrm{el}}\approx\tfrac12 E\varepsilon_{110}^2$;
for the strain magnitudes observed here ($\varepsilon_{110}\sim10^{-3}$) this
amounts to roughly $10^{-1}~\si{\mega\joule\per\meter\cubed}$. The second is the line energy of the stored dislocation content,
$u_{\rho}\approx\alpha G b^{2}\rho$, where $G$ is the shear modulus,
$b$ is the Burgers vector, $\rho$ is the dislocation density, and
$\alpha\approx0.5$ is a numerical prefactor. At the dislocation densities
characteristic of a 50\,\% cold reduction in Fe--3\,\%\,Si,
$\rho\sim10^{14}~\si{\per\meter\squared}$, and adopting
$G=82~\si{\giga\pascal}$ ~\cite{10.1063/1.1661710} and $b=2.48~\si{\angstrom}$, one obtains
$u_{\rho}\approx0.25~\si{\mega\joule\per\meter\cubed}$.
The third is the interfacial energy of the GNBs themselves, associated with their
misorientation~\cite{PhysRev.78.275}. The GNBs and the
grain boundary may be viewed as the local hotspots of the deformed state not because of any
single field but because they are simultaneously high in all three
reservoirs: net GND content, an adjacent dense dislocation population, and
interfacial energy. Recovery should therefore initiate preferentially at the
GNBs, where dislocation rearrangement and annihilation release the largest
energy per unit volume and where the long-range elastic strain can relax, in
agreement with the classical observation that subgrain coarsening and
nucleation begin at the highest-misorientation
boundaries~\cite{humphreys1995recrystallization,DOHERTY199739}.
Recrystallization nuclei are likewise expected at the most strongly
misoriented GNBs and at the grain boundaries, where orientation gradient,
stored energy, and boundary mobility together supply both the driving force
and the mobile interface. The three-dimensional misorientation and
strain-gradient fields reported here thus provide spatially resolved,
bulk-representative initial conditions for mesoscale models of
recrystallization in ferritic materials such as electrical steels, in which the final texture, and with
it the magnetic performance, is inherited from the nucleation landscape of
the deformed state~\cite{Hutchinson1999,Kestens01092016}. In Fe-Si alloys the stored energy is not distributed uniformly across different crystallographic orientations:
the principal rolling-texture fibers, in particular the $\gamma$-fiber
\(\langle111\rangle \parallel \mathrm{ND}\)) and the $\alpha$-fiber (\(\langle110\rangle \parallel \mathrm{RD}\)) ] \cite{Kestens01092016}, accumulate different dislocation contents and therefore different
stored energies, and it is this difference that sets the sequence in which
they recrystallize and hence the final texture that governs core loss and
permeability~\cite{doi:10.1080/02670836.2016.1231746,MAVRIKAKIS201992}. 
 
\section{Conclusions}\label{sec:conclusions}

In this work, we mapped the dislocation substructure and the residual
elastic strain field in three dimensions within a single grain in the bulk of
a 50\% cold-rolled Fe~3\%Si sample. The principal conclusions are:

\begin{enumerate}

\item Multi-peak analysis of DFXM data resolves GNBs and IDB dislocation cell structures
non-destructively in three dimensions at an industrially relevant
deformation level. The 3D-segmented cell size distribution and KAM
statistics agree with TEM literature values, confirming the representativeness of the
approach ~\cite{Shukla2026DislocationCells}.

\item Adjacent GNB-separated subdomains carry distinct mean residual
strain levels. The strain within each subdomain is comparatively
homogeneous, providing a direct 3D image of long range intragranular strain
partitioning. In contrast, the short range strain field , is relatively homogenous.

\item Correlating the dislocation (plastic) substructure with the residual elastic strain field
shows that GNBs accommodate nearly all the long-range residual elastic
strain, whereas plastic slip propagates into the GNB-bounded interiors
to form IDB cells with relatively more homogeneous strain values, direct
experimental separation of the elastic and plastic accommodation roles
of the two boundary classes.

\item Misorientation gradients (\si{\degree\per\micro\meter}) quantified along ND and TD experimentally show that GNBs generate a misorientation across a specific crystallographic direction and hence maybe invisible in DFXM COM maps depending on angle of the misorientation axis with the rotation axis of the DFXM scan. 

\section{Outlook}\label{sec:outlook}

The current study can be improved in several ways. In its current
configuration DFXM measures a single elastic strain component along one
scattering vector; recovery of the full strain tensor, and with it the
separation of the hydrostatic and deviatoric parts that would distinguish a
triaxially constrained boundary from a uniaxial one. This requires
multi-reflection measurements that are feasible but considerably more
demanding in beam time~\cite{poulsen, detlefs2025oblique}. The selection of the
strongest local intensity maximum as the representative orientation of each
gauge volume may bias the maps towards the larger or less deformed
cells~\cite{Shukla2026DislocationCells}, and the rigorous treatment of
multiple orientation contributions per gauge volume remains an active area
of method development~\cite{Henningsson2026Darling}. Finally, the analysis
pertains to a single grain of a single texture component; extending it to
several grains spanning the $\alpha$- and $\gamma$-fibers will be required
before the boundary-scale strain partitioning established here can be
connected to texture-level recrystallization statistics. Recently-developed synchrotron technique scanning 3DXRD can offer solutions for this \cite{henningsson2024microstructure}
 
The workflow presented here opens several promising experimental directions. In-situ annealing of the grain volume mapped in this study will enable direct imaging of elastic strain release and boundary migration and annihilation during recovery, and thereby providing a real time test of theoretical models of preferential nucleation at high energy GNBs. Combining the curvature-derived GND density with a conventional XRD line-broadening analysis of the total dislocation density would further yield the three-dimensional GND/SSD partition. Together, the measured orientation and strain fields can serve simultaneously as initial conditions and validation data for crystal plasticity and phase-field models, closing the long-standing loop between bulk experiment and mesoscale simulation in the predictive modeling of recrystallization in metals deformed at industrially-relevant levels.

\end{enumerate}

\backmatter

\section*{Acknowledgements}
We acknowledge the provision of beamtime MA6232 (Long Term Project) at
beamline ID03 of the European Synchrotron Radiation Facility (ESRF),
Grenoble, France. A.S.\ and C.Y.\ acknowledge ERC funding within the
D-REX project, grant no.\ 101116911.

\section*{Data Availibility}
The raw EBSD and DFXM data are available from
        the corresponding author upon reasonable request.

        \section*{Author Contributions}
C.Y.\ designed and performed the
        DFXM experiments. N.M.\ provided and processed the Fe~3\%Si
        material, and conducted the EBSD experiments. A.S.\ designed and performed the
        DFXM experiments,  data analysis, developed the
        segmentation workflow, and wrote the manuscript together with C.Y. All authors
        discussed the results and reviewed the manuscript.
        
\section*{Declarations}
The authors declare no competing
        interests.

\makeatletter
\patchcmd{\thebibliography}
  {\settowidth\labelwidth{\@biblabel{#1}}}
  {\settowidth\labelwidth{\@biblabel{#1}}\itemsep 0pt \parsep 0pt \topsep 0pt}
  {}{}
\makeatother
{\footnotesize

\bibliography{bib}

@article{WRIGHT2020100818,
title = {New opportunities at the Materials Science Beamline at ESRF to exploit high energy nano-focus X-ray beams},
journal = {Current Opinion in Solid State and Materials Science},
volume = {24},
number = {2},
pages = {100818},
year = {2020},
issn = {1359-0286},
author = {Jonathan Wright and Carlotta Giacobbe and Marta Majkut},
}

@article{yildirim-2025,
	author = {Yildirim, Can and Shukla, Aditya and Zhang, Yubin and Mavrikakis, Nikolas and Lesage, Louis and Sanna, Virginia and Sarkis, Marilyn and Li, Yaozhu and La Bella, Michela and Detlefs, Carsten and Poulsen, Henning Friis},
	journal = {Commun. Mater.},
	month = {8},
	number = {1},
	pages = {198},
	title = {{3D/4D imaging of complex and deformed microstructures with pink-beam dark field X-ray microscopy}},
	volume = {6},
	year = {2025},
	doi = {10.1038/s43246-025-00926-9},
}

@article{poulsen,
	author = {Poulsen, H. F. and Jakobsen, A. C. and Simons, H. and Ahl, S. R. and Cook, P. K. and Detlefs, C.},
	journal = {J. Appl. Crystallogr.},
	month = {9},
	number = {5},
	pages = {1441--1456},
	title = {{X-ray diffraction microscopy based on refractive optics}},
	volume = {50},
	year = {2017},
	doi = {10.1107/s1600576717011037},
}

@article{yildirim-2022,
	author = {Yildirim, C. and Mavrikakis, N. and Cook, P.K. and Rodriguez-Lamas, R. and Kutsal, M. and Poulsen, H.F. and Detlefs, C.},
	journal = {Scr. Mater.},
	month = {3},
	pages = {114689},
	title = {{4D microstructural evolution in a heavily deformed ferritic alloy: A new perspective in recrystallisation studies}},
	volume = {214},
	year = {2022},
	doi = {10.1016/j.scriptamat.2022.114689},
}

@misc{shukla_bridging_2025,
	doi = {10.48550/arXiv.2508.17897},
	publisher = {arXiv},
	author = {Shukla, Aditya and Yildirim, Can and Ball, James A. D. and Detlefs, Carsten and Cretton, Adam A. W. and Sarkis, Marilyn and Bella, Michela La and Ludwig, Wolfgang and Zhang, Yubin and Henningsson, Nils Axel},
	month = aug,
    title = {Bridging Grain Mapping and Dark Field X-ray Microscopy for Multiscale Diffraction Imaging},
	year = {2025},
	note = {[arXiv:2508.17897]},
}

@article{id03,
	author = {Isern, H. and Brochard, T. and Dufrane, T. and Brumund, P. and Papillon, E. and Scortani, D. and Hino, R. and Yildirim, C. and Lamas, R. Rodriguez and Li, Y. and Sarkis, M. and Detlefs, C.},
	journal = {J. Phys.: Conf. Ser.},
	month = {5},
	number = {1},
	pages = {012163},
	title = {{The ESRF dark-field x-ray microscope at ID03}},
	volume = {3010},
	year = {2025},
	doi = {10.1088/1742-6596/3010/1/012163},
}

@article{simons-2015,
	author = {Simons, H. and King, A. and Ludwig, W. and Detlefs, C. and Pantleon, W. and Schmidt, S. and Stöhr, F. and Snigireva, I. and Snigirev, A. and Poulsen, H. F.},
	journal = {Nat. Commun.},
	month = {1},
	number = {1},
	pages = {6098},
	title = {{Dark-field X-ray microscopy for multiscale structural characterization}},
	volume = {6},
	year = {2015},
	doi = {10.1038/ncomms7098},
}

@article{darfix,
	author = {Ferrer, Júlia Garriga and Rodríguez-Lamas, Raquel and Payno, Henri and De Nolf, Wout and Cook, Phil and Jover, Vicente Armando Solé and Yildirim, Can and Detlefs, Carsten},
	journal = {J. Synchrotron Radiat.},
	month = {3},
	number = {3},
	pages = {527--537},
	title = {{darfix – data analysis for dark-field X-ray microscopy}},
	volume = {30},
	year = {2023},
	doi = {10.1107/s1600577523001674},
}

@article{chen_cells_2004,
	title = {On cells and microbands formed in an interstitial-free steel during cold rolling at low to medium reductions},
	volume = {35},
	doi = {10.1007/s11661-004-0178-5},
	language = {en},
	number = {11},
	urldate = {2026-05-01},
	journal = {Metall. Mater. Trans. A},
	author = {Chen, Q. Z. and Duggan, B. J.},
	month = nov,
	year = {2004},
	pages = {3423--3430},
}

@misc{darling,
  author       = {Henningsson, Axel},
  title        = {Darling: open source Python package for DFXM reconstruction},
  year         = {2026},
  url          = {https://github.com/AxelHenningsson/darling},
  note         = {Accessed: 2026-05-01}
}

@article{zelenika202,
  title={Observing formation and evolution of dislocation cells during plastic deformation},
  author={Zelenika, Albert and Cretton, Adam Andr{\'e} William and Frankus, Felix and Borgi, Sina and Grumsen, Flemming B and Yildirim, Can and Detlefs, Carsten and Winther, Grethe and Poulsen, Henning Friis},
  journal = {Sci. Rep.},
  volume={15},
  number={1},
  pages={8655},
  year={2025},
  publisher={Nature Publishing Group UK London}
}

@article{adams_seeded_1994,
	title = {Seeded region growing},
	volume = {16},
	copyright = {https://ieeexplore.ieee.org/Xplorehelp/downloads/license-information/IEEE.html},
	doi = {10.1109/34.295913},
	number = {6},
	urldate = {2026-05-01},
	journal = {IEEE Trans. Pattern Anal. Mach. Intell.},
	author = {Adams, R. and Bischof, L.},
	month = jun,
	year = {1994},
	pages = {641--647},
}

@article{LI20041069,
title = {Microstructural evolution of IF-steel during cold rolling},
journal = {Acta Mater.},
volume = {52},
number = {4},
pages = {1069-1081},
year = {2004},
doi = {https://doi.org/10.1016/j.actamat.2003.10.040},
author = {B.L Li and A Godfrey and Q.C Meng and Q Liu and N Hansen}}

@article{HUGHES2003147,
title = {Geometrically necessary boundaries, incidental dislocation boundaries and geometrically necessary dislocations},
journal = {Scr. Mater.},
volume = {48},
number = {2},
pages = {147-153},
year = {2003},
doi = {https://doi.org/10.1016/S1359-6462(02)00358-5},

author = {D.A Hughes and N Hansen and D.J Bammann}
}

@article{LIU19985819,
title = {Effect of grain orientation on deformation structure in cold-rolled polycrystalline aluminium},
journal = {Acta Mater.},
volume = {46},
number = {16},
year = {1998},
pages = {5819-5838},
doi = {https://doi.org/10.1016/S1359-6454(98)00229-8},

author = {Q. Liu and D. {Juul Jensen} and N. Hansen}}

@article{Hansen01082011,
author = {N Hansen and D Juul Jensen},
title = {Deformed metals – structure, recrystallisation and strength},
journal = {Mater. Sci. Technol.},
volume = {27},
number = {8},
pages = {1229--1240},
year = {2011},
publisher = {Taylor \& Francis}}

@article{YU20136577,
title = {Linking recovery and recrystallization through triple junction motion in aluminum cold rolled to a large strain},
journal = {Acta Mater.},
volume = {61},
number = {17},
pages = {6577-6586},
year = {2013},
author = {Tianbo Yu and Niels Hansen and Xiaoxu Huang},}

@article{RAABE2002421,
title = {Theory of orientation gradients in plastically strained crystals},
journal = {Acta Mater.},
volume = {50},
number = {2},
pages = {421-440},
year = {2002},
doi = {https://doi.org/10.1016/S1359-6454(01)00323-8},

author = {D Raabe and Z Zhao and S.-J Park and F Roters}
}

@article{Henningsson2026Darling,
  author  = {Henningsson, Axel and Frankus, Felix Tristan and Cretton, Adam A. W. and Shukla, Aditya and Gayoso Padula, Antonella and La Bella, Michela and Haack, Johann and Staeck, Steffen and Poulsen, Henning Friis and Winther, Grethe},
  title   = {Multi-peak diffraction analysis for enhanced orientation mapping in dark-field X-ray microscopy of deformed metals using Darling},
  journal = {IOP Conference Series: Materials Science and Engineering},
  year    = {2026},
  note    = {Submitted to the proceedings of the 46th Ris{\o} International Symposium on Materials Science: Characterization of Evolving Microstructures in Metals. In press.}
}

@article{henningsson2024microstructure,
  title={Microstructure and stress mapping in 3D at industrially relevant degrees of plastic deformation},
  author={Henningsson, Axel and Kutsal, Mustafacan and Wright, Jonathan P and Ludwig, Wolfgang and S{\o}rensen, Henning Osholm and Hall, Stephen A and Winther, Grethe and Poulsen, Henning Friis},
  journal={Scientific Reports},
  volume={14},
  number={1},
  pages={20213},
  year={2024},
  publisher={Nature Publishing Group UK London}
}

@article{detlefs2025oblique,
  title={Oblique diffraction geometry for the observation of several non-coplanar Bragg reflections under identical illumination},
  author={Detlefs, Carsten and Henningsson, Axel and Kanesalingam, Brinthan and Cretton, Adam AW and Corley-Wiciak, Cedric and Frankus, Felix T and Pal, Dayeeta and Irvine, Sara and Borgi, Sina and Poulsen, Henning F and others},
  journal={Applied Crystallography},
  volume={58},
  number={4},
  year={2025},
  publisher={International Union of Crystallography}
}

@article{Shukla2026DislocationCells,
  author  = {Shukla, Aditya and Henningsson, Axel and Cretton, Adam A. W. and Yildirim, Can},
  title   = {Direct visualization, segmentation and quantification of dislocation cells in cold-rolled Fe-3\% Si using dark-field X-ray microscopy},
  journal = {IOP Conference Series: Materials Science and Engineering},
  year    = {2026},
  note    = {Submitted to the proceedings of the 46th Ris{\o} International Symposium on Materials Science: Characterization of Evolving Microstructures in Metals. In press.}
}

@article{BEVER19735,
title = {The stored energy of cold work},
journal = {Progress in Materials Science},
volume = {17},
pages = {5-177},
year = {1973},
issn = {0079-6425},
doi = {https://doi.org/10.1016/0079-6425(73)90001-7},
url = {https://www.sciencedirect.com/science/article/pii/0079642573900017},
author = {M.B. Bever and D.L. Holt and A.L. Titchener}
}

@article{NYE1953153,
title = {Some geometrical relations in dislocated crystals},
journal = {Acta Metallurgica},
volume = {1},
number = {2},
pages = {153-162},
year = {1953},
issn = {0001-6160},
doi = {https://doi.org/10.1016/0001-6160(53)90054-6},
url = {https://www.sciencedirect.com/science/article/pii/0001616053900546},
author = {J.F Nye}
}

@article{Ashby01021970,
author = {M. F. Ashby},
title = {The deformation of plastically non-homogeneous materials},
journal = {The Philosophical Magazine: A Journal of Theoretical Experimental and Applied Physics},
volume = {21},
number = {170},
pages = {399--424},
year = {1970},
publisher = {Taylor \& Francis},
doi = {10.1080/14786437008238426},


URL = { 
https://doi.org/10.1080/14786437008238426

},
eprint = { 
 https://doi.org/10.1080/14786437008238426
}

}

@book{humphreys1995recrystallization,
  title     = {Recrystallization and Related Annealing Phenomena},
  author    = {Humphreys, F. J. and Hatherly, M.},
  year      = {1995},
  publisher = {Pergamon},
  address   = {Oxford, U.K.},
  isbn      = {978-0-08-041884-3}
}

@article{doi:10.1080/02670836.2016.1231746,
author = {L. A. I. Kestens and H. Pirgazi},
title ={Texture formation in metal alloys with cubic crystal structures},

journal = {Materials Science and Technology},
volume = {32},
number = {13},
pages = {1303-1315},
year = {2016},
doi = {10.1080/02670836.2016.1231746},

URL = { 
    
        https://doi.org/10.1080/02670836.2016.1231746
    
    

},
eprint = { 
    
        https://doi.org/10.1080/02670836.2016.1231746
    
    

}}

@article{Hutchinson1999,
  author  = {Hutchinson, B.},
  title   = {Deformation microstructures and textures in steels},
  journal = {Philosophical Transactions of the Royal Society A},
  year    = {1999},
  volume  = {357},
  number  = {1756},
  pages   = {1471--1485},
  doi     = {10.1098/rsta.1999.0385}
}

@article{WILKINSON2012366,
title = {Strains, planes, and EBSD in materials science},
journal = {Materials Today},
volume = {15},
number = {9},
pages = {366-376},
year = {2012},
issn = {1369-7021},
doi = {https://doi.org/10.1016/S1369-7021(12)70163-3},
url = {https://www.sciencedirect.com/science/article/pii/S1369702112701633},
author = {Angus J. Wilkinson and T. Ben. Britton}
}

@article{yildirim_probing_2020,
	title = {Probing nanoscale structure and strain by dark-field x-ray microscopy},
	volume = {45},
	issn = {1938-1425},
	url = {https://doi.org/10.1557/mrs.2020.89},
	doi = {10.1557/mrs.2020.89},
	language = {en},
	number = {4},
	urldate = {2026-06-22},
	journal = {MRS Bulletin},
	author = {Yildirim, Can and Cook, Phil and Detlefs, Carsten and Simons, Hugh and Poulsen, Henning Friis},
	month = apr,
	year = {2020},
	pages = {277--282},
}

@article{MAVRIKAKIS201992,
title = {A multi-scale study of the interaction of Sn solutes with dislocations during static recovery in a-Fe},
journal = {Acta Materialia},
volume = {174},
pages = {92-104},
year = {2019},
issn = {1359-6454},
doi = {https://doi.org/10.1016/j.actamat.2019.05.021},
url = {https://www.sciencedirect.com/science/article/pii/S1359645419303039},
author = {N. Mavrikakis and C. Detlefs and P.K. Cook and M. Kutsal and A.P.C. Campos and M. Gauvin and P.R. Calvillo and W. Saikaly and R. Hubert and H.F. Poulsen and A. Vaugeois and H. Zapolsky and D. Mangelinck and M. Dumont and C. Yildirim}
}

@article{GODFREY20001897,
title = {Scaling of the spacing of deformation induced dislocation boundaries},
journal = {Acta Materialia},
volume = {48},
number = {8},
pages = {1897-1905},
year = {2000},
issn = {1359-6454},
doi = {https://doi.org/10.1016/S1359-6454(99)00474-7},
url = {https://www.sciencedirect.com/science/article/pii/S1359645499004747},
author = {A Godfrey and D.A Hughes}
}

@article{Wright2011,
  author  = {Wright, S. I. and Nowell, M. M. and Field, D. P.},
  title   = {A Review of Strain Analysis Using Electron Backscatter 
             Diffraction},
  journal = {Microscopy and Microanalysis},
  year    = {2011},
  volume  = {17},
  pages   = {316--329},
  doi     = {10.1017/S1431927611000055}
}

@article{Pantleon2008,
  author  = {Pantleon, W.},
  title   = {Resolving the geometrically necessary dislocation content 
             by conventional electron backscattering diffraction},
  journal = {Scripta Materialia},
  year    = {2008},
  volume  = {58},
  pages   = {994--997},
  doi     = {10.1016/j.scriptamat.2008.01.050}
}

@article{KAMAYA201256,
title = {Assessment of local deformation using EBSD: Quantification of local damage at grain boundaries},
journal = {Materials Characterization},
volume = {66},
pages = {56-67},
year = {2012},
issn = {1044-5803},
doi = {https://doi.org/10.1016/j.matchar.2012.02.001},
url = {https://www.sciencedirect.com/science/article/pii/S1044580312000265},
author = {Masayuki Kamaya}
}

@ARTICLE{Sanna2026-vj,
  title         = "{3D} mapping of intragranular residual strain and
                   microstructure in recrystallized iron using dark-field X-ray
                   microscopy",
  author        = "Sanna, Virginia and Zhang, Yubin and Ludwig, Wolfgang and
                   Shukla, Aditya and Benhadjira, Abderrahmane and Sarkis,
                   Marilyn and Yildirim, Can",
  month         =  mar,
  year          =  2026,
  copyright     = "http://arxiv.org/licenses/nonexclusive-distrib/1.0/",
  archivePrefix = "arXiv",
  primaryClass  = "cond-mat.mtrl-sci",
  eprint        = "2603.08968"
}

@Article{met14101127,
AUTHOR = {Gao, Yijing and Xu, Yunbo and Chen, Haoran and Yuan, Bingyu and Gao, Zhenyu and Zhou, Lifeng},
TITLE = {Dislocation Strengthening and Texture Evolution of Non-Oriented Fe-3.3 wt% Si Steel in Double Cold Rolling},
JOURNAL = {Metals},
VOLUME = {14},
YEAR = {2024},
NUMBER = {10},
ARTICLE-NUMBER = {1127},
URL = {https://www.mdpi.com/2075-4701/14/10/1127},
ISSN = {2075-4701},
DOI = {10.3390/met14101127}
}

@article{PhysRev.78.275,
  title = {Dislocation Models of Crystal Grain Boundaries},
  author = {Read, W. T. and Shockley, W.},
  journal = {Phys. Rev.},
  volume = {78},
  issue = {3},
  pages = {275--289},
  numpages = {0},
  year = {1950},
  month = {May},
  publisher = {American Physical Society},
  doi = {10.1103/PhysRev.78.275},
  url = {https://link.aps.org/doi/10.1103/PhysRev.78.275}
}

@article{KASSNER201344,
title = {Long range internal stresses in single-phase crystalline materials},
journal = {International Journal of Plasticity},
volume = {45},
pages = {44-60},
year = {2013},
note = {In Honor of Rob Wagoner},
issn = {0749-6419},
doi = {https://doi.org/10.1016/j.ijplas.2012.10.003},
url = {https://www.sciencedirect.com/science/article/pii/S0749641912001568},
author = {M.E. Kassner and P. Geantil and L.E. Levine}
}

@article{DOHERTY199739,
title = {Recrystallization and texture},
journal = {Progress in Materials Science},
volume = {42},
number = {1},
pages = {39-58},
year = {1997},
issn = {0079-6425},
doi = {https://doi.org/10.1016/S0079-6425(97)00007-8},
url = {https://www.sciencedirect.com/science/article/pii/S0079642597000078},
author = {Roger D. Doherty}
}

@article{Kestens01092016,
author = {L. A. I. Kestens and H. Pirgazi},
title = {Texture formation in metal alloys with cubic crystal structures},
journal = {Materials Science and Technology},
volume = {32},
number = {13},
pages = {1303--1315},
year = {2016},
publisher = {Taylor \& Francis},
doi = {10.1080/02670836.2016.1231746},


URL = { 
    
        https://doi.org/10.1080/02670836.2016.1231746
    
    

},
eprint = { 
    
        https://doi.org/10.1080/02670836.2016.1231746
    
    

}

}

@article{MUGHRABI19831367,
title = {Dislocation wall and cell structures and long-range internal stresses in deformed metal crystals},
journal = {Acta Metallurgica},
volume = {31},
number = {9},
pages = {1367-1379},
year = {1983},
issn = {0001-6160},
doi = {https://doi.org/10.1016/0001-6160(83)90007-X},
url = {https://www.sciencedirect.com/science/article/pii/000161608390007X},
author = {H. Mughrabi}
}

@article{10.1063/1.1661710,
    author = {Dever, D.J.},
    title = {Temperature dependence of the elastic constants in a‐iron single crystals: relationship to spin order and diffusion anomalies},
    journal = {Journal of Applied Physics},
    volume = {43},
    number = {8},
    pages = {3293-3301},
    year = {1972},
    month = {08},
    issn = {0021-8979},
    doi = {10.1063/1.1661710},
}
}

\end{document}